\documentclass[reprint, superscriptaddress, secnumarabic, amssymb, nobibnotes, aps, prl]{revtex4-1}

\usepackage{graphicx}
\usepackage{epstopdf}
\usepackage[T1]{fontenc}
\usepackage[latin9]{inputenc}
\usepackage{amsbsy}
\usepackage{gensymb}
\usepackage[T1]{fontenc}
\usepackage[latin9]{inputenc}
\usepackage{amsmath}
\usepackage{amssymb}
\usepackage{bbm}
\usepackage{braket}
\usepackage{xcolor}
\allowdisplaybreaks
\usepackage{graphicx}
\usepackage[colorlinks=true]{hyperref}  
\hypersetup{
    bookmarks=true,         
    unicode=false,          
    pdftoolbar=true,        
    pdfmenubar=true,        
    pdffitwindow=false,     
    pdfstartview={FitH},    
    pdftitle={x},    
    pdfauthor={SM},     
    pdfsubject={},   
    pdfcreator={},   
    pdfproducer={}, 
    pdfkeywords={} {} {}, 
    pdfnewwindow=true,      
    colorlinks=true,       
    linkcolor=blue, 
    citecolor=blue,        
    filecolor=magenta,      
    urlcolor=blue           
} 
\usepackage[normalem]{ulem}

\renewcommand{\approx}{\simeq}

\begin{document}
\title{\textrm{Structural, magnetic and x-ray absorption spectroscopy studies of new Cr-based low, medium and high-entropy spinel oxides}}

\author{Sushanta Mandal}
\affiliation{Department of Physics and Materials Science, Thapar Institute of Engineering and Technology, Patiala 147004, India}

\author{Jyoti Sharma}
\affiliation{UGC-DAE Consortium for Scientific Research, Khandwa Road, Indore 452001, Madhya Pradesh, India}
\author{Tirthankar Chakraborty}
\affiliation{Department of Physics and Materials Science, Thapar Institute of Engineering and Technology, Patiala 147004, India}
\author{Sanjoy Kr. Mahatha}
\affiliation{UGC-DAE Consortium for Scientific Research, Khandwa Road, Indore 452001, Madhya Pradesh, India}
\author{Sourav Marik}
\email[]{soumarik@thapar.edu}
\affiliation{Department of Physics and Materials Science, Thapar Institute of Engineering and Technology, Patiala 147004, India}

\begin{abstract}
\begin{flushleft}

\end{flushleft}
The emergence of high-entropy oxides has spurred significant research interest in recent times. These compounds exhibit exotic functional properties that often transcend simple linear combinations of their constituent elements. Herein, we present a new series of Cr-based low, medium, and high entropy spinel oxides with composition NiCr$_{2}$O$_{4}$, [Ni$_{0.5}$Mn$_{0.5}$]Cr$_{2}$O$_{4}$, [Ni$_{0.33}$Mn$_{0.33}$Co$_{0.33}$]Cr$_{2}$O$_{4}$, [Ni$_{0.25}$Mn$_{0.25}$Co$_{0.25}$Cu$_{0.25}$]Cr$_{2}$O$_{4}$, [Ni$_{0.2}$Mn$_{0.2}$Co$_{0.2}$Cu$_{0.2}$Zn$_{0.2}$]Cr$_{2}$O$_{4}$, and [Ni$_{0.2}$Mg$_{0.2}$Co$_{0.2}$Cu$_{0.2}$Zn$_{0.2}$]Cr$_{2}$O$_{4}$. We conducted detailed structural (X-ray and Neutron diffraction), microstructural, Raman spectroscopy, magnetic, and X-ray absorption spectroscopy measurements on these materials. Our study reveals that the incorporation of multiple cations at the A-site of the structure (AB$_{2}$O$_{4}$) significantly modulates the magnetic properties. These compounds exhibit transitions from complex ferrimagnetic ([Ni$_{0.2}$Mn$_{0.2}$Co$_{0.2}$Cu$_{0.2}$Zn$_{0.2}$]Cr$_{2}$O$_{4}$) to antiferromagnetic ([Ni$_{0.2}$Mg$_{0.2}$Co$_{0.2}$Cu$_{0.2}$Zn$_{0.2}$]Cr$_{2}$O$_{4}$) states, with remarkable coercivity variations, demonstrating the ability to tailor magnetic responses through compositional design. 
\end{abstract}
\maketitle
\section{Introduction}
Recently, high entropy oxides (HEOs) have emerged as an exciting and promising field of research, fundamentally reshaping material design and properties \cite{1, 2}. The configurational entropy ($\Delta$S$_{conf.}$) is greatly enhanced by randomly incorporating multiple elements (five or more in an equimolar ratio) into a single crystallographic site. This highly enhanced configurational entropy can stabilize the structure of a compound, leading to what is termed an entropy-stabilized phase.

Rost et al. in 2015, stabilized and investigated the first high entropy oxide (HEO) having composition (Mg$_{0.2}$Co$_{0.2}$Ni$_{0.2}$Cu$_{0.2}$Zn$_{0.2}$)O \cite{3}. Subsequently, high entropy oxides (HEOs) have attracted considerable research interest due to their facile synthesis, outstanding stability, extraordinary functional properties, and the augmented design versatility offered by the possibility of the presence of multiple cations in the structure \cite{1,2,3,4}.  Illustrative examples include tunable magnetism \cite{5}, high exchange bias effect \cite{6}, and extremely high magnetic frustration \cite{7} in high entropy spinel oxides, spin reorientation in high entropy perovskites \cite{8}, robust ferrimagnetism in (CrMnFeCoNi)$_{3}$O$_{4}$ film \cite{9}, extremely high value of colossal magnetoresistance in strongly correlated high-entropy manganites \cite{10}, enhanced catalytic performance in Al-based high entropy spinels \cite{11}, enhanced corrosion resistance in high entropy pyrochlores \cite{12}, promising thermoelectric properties with ultralow (record low) thermal conductivity in high entropy oxides with a tungsten bronze structure \cite{13} and enhanced thermal radiation \cite{14} in high entropy type spinel oxides. The increased disorder within entropy-stabilized systems brings significant enhancements in their functional properties. High entropy stabilization offers a promising framework for finely tuning materials at the atomic scale, enabling the fabrication of materials with tailored properties to address specific needs. However, synthesizing high entropy oxides (HEOs) with precise compositions and desired properties poses a significant challenge. A precise selection of constituent elements and their stoichiometric ratios makes it feasible to engineer materials with the desired characteristics suitable for targeted applications. Thus, a compelling motivation exists to explore and investigate new strongly correlated high entropy oxides having novel structures and intriguing properties. Herein, we present a new series of Cr-based low, medium, and high entropy spinel oxides. We have highlighted the detailed and systematic structural, magnetic, Raman spectroscopy, and X-ray absorption spectroscopy studies on these new compounds.

Spinel oxide system stands as a fascinating category of material system, not only for their broad utility across diverse applications but also for their plethora of novel and intriguing physicochemical properties. These include phenomena such as frustrated magnetism, multiferroic behavior, orbital glass states, and the emergence of spin-orbital liquids \cite{15,16,17,18}. Structurally, they adhere to a distinct framework characterized by the general formula AB$_2$O$_4$, with A and B representing metal ions. Within this structural framework, metal cations are both in octahedral and tetrahedral coordination environments, surrounded by oxygen atoms, leading to the formation of two distinct sets of magnetic sublattices. Conventionally, B cations occupy octahedral sites, configuring a lattice akin to pyrochlore and thereby inducing highly frustrated magnetic interactions \cite{19}. The geometrical frustration within the pyrochlore lattice yields a plethora of intriguing physical effects, such as multiferroics, Spin-Peierls-like phase transitions, and quantum magnetic frustration \cite{15,16,17,18,20}. These effects emerge from the interplay between the lattice's structure and the orientation of its spins, highlighting the correlation between structural and spin degrees of freedom. Conversely, A ions occupy tetrahedral sites, forming an eightfold coordination reminiscent of a diamond lattice structure \cite{21}. This bipartite lattice arrangement can be envisaged as comprising two face-centered interpenetrating cubic (fcc) sublattices, each diagonally shifted by one-quarter of the lattice parameter. Variations in the composition of magnetic and non-magnetic cations across both tetrahedral A and octahedral B sites can give rise to intricate magnetic interactions, thus fostering the emergence of exotic properties and states such as spiral spin liquid phases \cite{22}, distinctive glassy magnetic behaviors \cite{23}, and spin-orbital liquids \cite{24} in spinel materials. The configuration of the pyrochlore lattice presents a rich and fertile landscape for both theoretical exploration and experimental investigation.

\begin{figure}
\includegraphics[width=1.1\columnwidth]{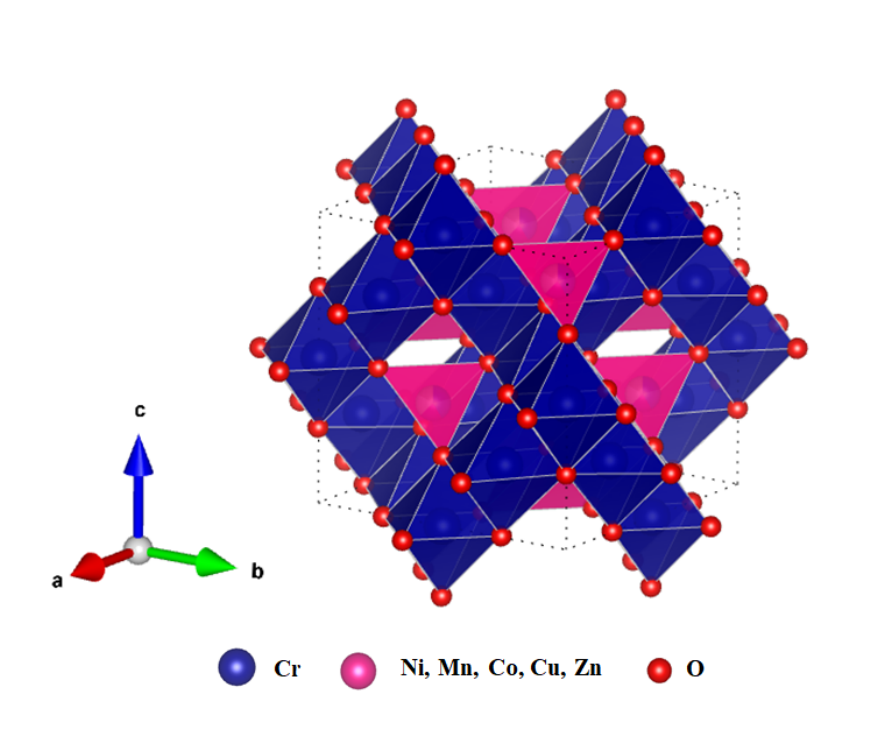}
\caption{\label{Figure 1:STR}Crystal structure of the spinel AB$_2$O$_4$ oxides. B cations (here Cr) occupy the octahedral sites, A cations (Ni, Mn, Mg, Co, Cu, Zn) occupy the tetrahedral sites}
\end{figure}

In the present paper, we have synthesized a new series of Cr-based low, medium, and high entropy spinel oxides having compositions NiCr$_{2}$O$_{4}$, [Ni$_{0.5}$Mn$_{0.5}$]Cr$_{2}$O$_{4}$, [Ni$_{0.33}$Mn$_{0.33}$Co$_{0.33}$]Cr$_{2}$O$_{4}$, [Ni$_{0.25}$Mn$_{0.25}$Co$_{0.25}$Cu$_{0.25}$]Cr$_{2}$O$_{4}$, [Ni$_{0.2}$Mn$_{0.2}$Co$_{0.2}$Cu$_{0.2}$Zn$_{0.2}$]Cr$_{2}$O$_{4}$, and [Ni$_{0.2}$Mg$_{0.2}$Co$_{0.2}$Cu$_{0.2}$Zn$_{0.2}$]Cr$_{2}$O$_{4}$ by systematically increasing the configuration entropy (and chemical complexity) in the A site of the Cr-based spinel structure. We have presented detailed and systematic structural (using X-ray powder diffraction, neutron powder diffraction, and electron microscopy), magnetic (magnetization), Raman spectroscopy, and X-ray absorption spectroscopy studies on these new compounds. A remarkable change in the magnetic properties (from ferrimagnetic to antiferromagnetic, unusually large coercivity) is observed with a small change in the A site of the structure. Our study shows the feasibility of accommodating numerous elements within a single sublattice of a complex spinel system and highlights prospects for tailoring and optimizing functional properties in high-entropy stabilized correlated electron systems.

\begin{figure*}
\includegraphics[width=2.0\columnwidth]{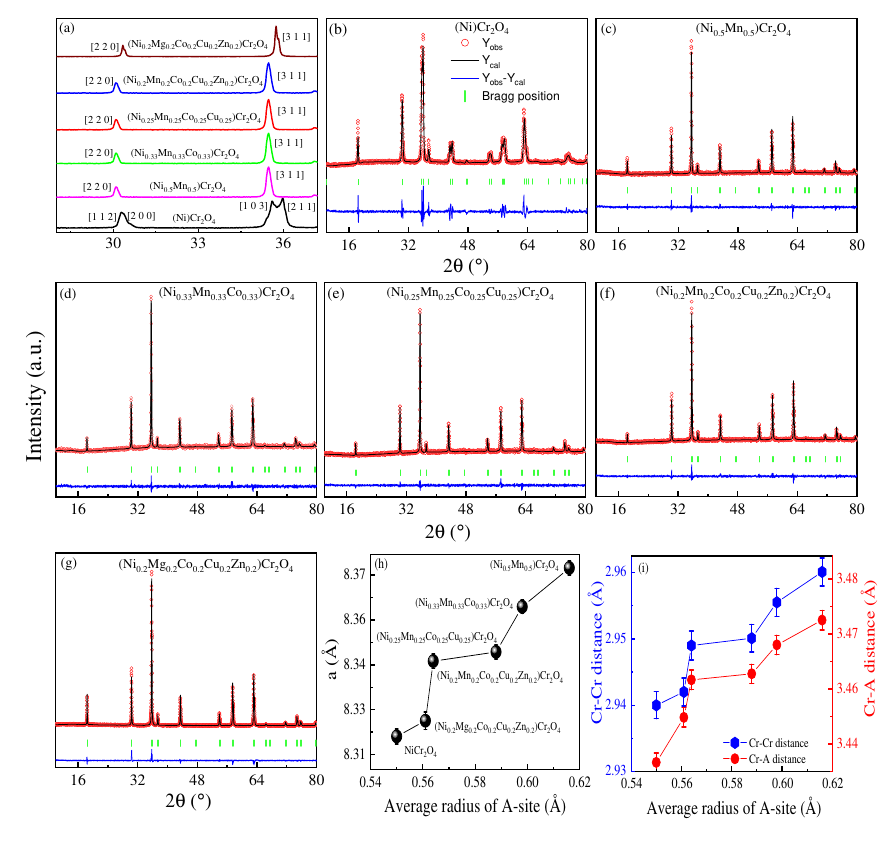}
\caption{\label{Figure 2:XRD} (a) Enlarged part of the room temperature X-ray diffraction (RT - XRD) patterns for NiCr$_{2}$O$_{4}$, [Ni$_{0.5}$Mn$_{0.5}$]Cr$_{2}$O$_{4}$, [Ni$_{0.33}$Mn$_{0.33}$Co$_{0.33}$]Cr$_{2}$O$_{4}$, [Ni$_{0.25}$Mn$_{0.25}$Co$_{0.25}$Cu$_{0.25}$]Cr$_{2}$O$_{4}$, [Ni$_{0.2}$Mn$_{0.2}$Co$_{0.2}$Cu$_{0.2}$Zn$_{0.2}$]Cr$_{2}$O$_{4}$, and [Ni$_{0.2}$Mg$_{0.2}$Co$_{0.2}$Cu$_{0.2}$Zn$_{0.2}$]Cr$_{2}$O$_{4}$. (b) - (g) Rietveld refinement plots of the RT - XRD patterns, variation of the (h) lattice parameters, and (i) Cr - Cr and Cr-A distances with the radius of the A site for all the samples. The data for the cubic NiCr$_{2}$O$_{4}$ is taken from Ref. \cite{26}.}
\end{figure*}
      
\section{Experimental Details}
\textbf{Synthesis.} The standard solid-state reaction route is used to prepare all the samples. Stoichiometric amounts of high-purity Co$_3$O$_4$ (99.9\%) MnO$_2$ (99.9\%), CuO (99.9\%), ZnO (99.9\%), NiO (99.9\%), MgO (99.9\%) and Cr$_2$O$_3$ (99.99\%) were employed. The precursor materials were thoroughly mixed using a mortar and pestle, followed by multiple heat treatments and air quenching. The final sintering was conducted at 1373 K for 36 hours.

\textbf{X-ray Diffraction and Neutron Diffraction.} X-ray diffraction (XRD) analysis was performed on the powdered samples using a PANalytical diffractometer with (Cu-K$_{\alpha}$, $\lambda$ = 1.54056 \text{\AA}) at ambient conditions. Neutron diffraction (NPD) data for the [Ni$_{0.2}$Mn$_{0.2}$Co$_{0.2}$Cu$_{0.2}$Zn$_{0.2}$]Cr$_{2}$O$_{4}$ and [Ni$_{0.2}$Mg$_{0.2}$Co$_{0.2}$Cu$_{0.2}$Zn$_{0.2}$]Cr$_{2}$O$_{4}$ were collected at ambient temperature at BARC, India ($\lambda$ = 1.48 \text{\AA}). We have performed the Rietveld refinements of the diffraction patterns (XRD and NPD) using the fullProf suite software. 

\textbf{Electron Microscopy.} Energy Dispersive X-ray Spectroscopy (EDS) mappings were obtained using a BRUKER XFlash 6160 system, while microstructure analysis was performed with a ZEISS GEMINI field emission scanning electron microscope (FE-SEM). High-angle annular dark-field scanning transmission electron microscopy (HAADF-STEM) was conducted using a JEOL ARM-200F microscope with cold FEG probe image aberration correction at 200 kV to investigate the crystal structure. All measurements were conducted at room temperature.

\begin{figure*}
\includegraphics[width=2.0\columnwidth]{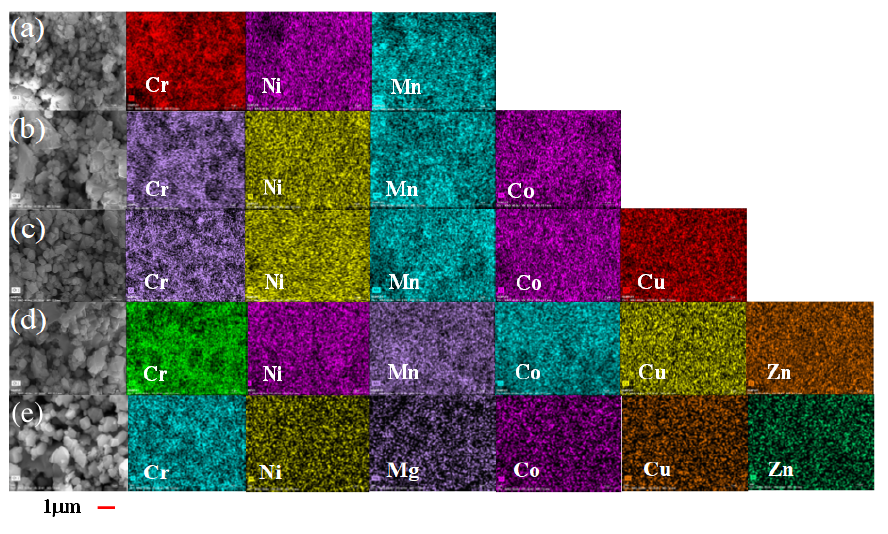}
\caption{\label{Figure 3:SEM} Representative FE-SEM images and the EDS elemental mapping results for (a) [Ni$_{0.5}$Mn$_{0.5}$]Cr$_{2}$O$_{4}$, (b) [Ni$_{0.33}$Mn$_{0.33}$Co$_{0.33}$]Cr$_{2}$O$_{4}$, (c) [Ni$_{0.25}$Mn$_{0.25}$Co$_{0.25}$Cu$_{0.25}$]Cr$_{2}$O$_{4}$, (d) [Ni$_{0.2}$Mn$_{0.2}$Co$_{0.2}$Cu$_{0.2}$Zn$_{0.2}$]Cr$_{2}$O$_{4}$, and (e) [Ni$_{0.2}$Mg$_{0.2}$Co$_{0.2}$Cu$_{0.2}$Zn$_{0.2}$]Cr$_{2}$O$_{4}$. All the materials show excellent homogeneity at the microscopic scale. The elemental ratios obtained from EDS elemental mapping are highlighted in Table I.}
\end{figure*} 

\begin{table*}
\small
  \caption{\ Compositions of all the materials based on EDS mapping analysis.}
  \label{tbl 1:EDS table}
  \begin{tabular*}{\textwidth}{@{\extracolsep{\fill}}ll}
    \hline
    Sample Compositions (stoichiometric) & Sample Compositions (EDS mapping)\\
    \hline
   [Ni$_{0.5}$Mn$_{0.5}$]Cr$_{2}$O$_{4}$ & [Ni$_{0.46}$Mn$_{0.55}$]Cr$_{1.99}$O$_{4}$\\

   [Ni$_{0.33}$Mn$_{0.33}$Co$_{0.33}$]Cr$_{2}$O$_{4}$ & [Ni$_{0.29}$Mn$_{0.27}$Co$_{0.45}$]Cr$_{1.99}$O$_{4}$\\
   
   [Ni$_{0.25}$Mn$_{0.25}$Co$_{0.25}$Cu$_{0.25}$]Cr$_{2}$O$_{4}$ & [Ni$_{0.23}$Mn$_{0.24}$Co$_{0.27}$Cu$_{0.25}$]Cr$_{2}$O$_{4}$\\

    [Ni$_{0.2}$Mn$_{0.2}$Co$_{0.2}$Cu$_{0.2}$Zn$_{0.2}$]Cr$_{2}$O$_{4}$ & [Ni$_{0.18}$Mn$_{0.23}$Co$_{0.17}$Cu$_{0.19}$Zn$_{0.17}$]Cr$_{2.04}$O$_{4}$\\

  [Ni$_{0.2}$Mg$_{0.2}$Co$_{0.2}$Cu$_{0.2}$Zn$_{0.2}$]Cr$_{2}$O$_{4}$ & [Ni$_{0.25}$Mg$_{0.09}$Co$_{0.23}$Cu$_{0.21}$Zn$_{0.21}$]Cr$_{2.03}$O$_{4}$\\
    \hline
  \end{tabular*}
\end{table*}

\textbf{Raman Spectra.} Raman spectra were collected from powdered samples under ambient conditions using a LabRAM HR Raman spectrometer (Horiba, France) with a 532 nm laser source.

\textbf{X-Ray Absorption Spectroscopy.} Soft X-ray absorption spectroscopy (SXAS) measurements were conducted at the Raja Ramanna Centre for Advanced Technology (RRCAT) in Indore, India. The SXAS beamline (BL-01) of the Indus-2 synchrotron source provided an energy resolution of approximately 0.5 eV for the measurements, all of which were performed at room temperature. 

\textbf{Magnetism.} Magnetization measurements, including temperature and magnetic field dependence, were performed using a Quantum Design MPMS 3 superconducting quantum interference device (SQUID). Both field-cooled (FC) and zero-field-cooled (ZFC) modes were utilized. The measurements were conducted over a temperature range of 4 K to 300 K.

\section{Results and Discussion}

Figure 2(a) highlights the enlarged part of the RT-XRD data for all the samples. The RT-XRD patterns confirm the phase purity of all the samples. As expected NiCr$_2$O$_4$ is found to crystallize in a tetragonal structure (space group I4$_1$/amd, structural transition temperature 310 K \cite{25, 26}) with lattice parameters of a = b = 5.8494 (1) {\AA} and c = 8.3773 (1) {\AA}. The Jahn-Teller active Ni$^{2+}$ in the tetrahedral site leads to the stabilization of a tetragonal structure. However, dilution in the Ni$^{2+}$ content in the tetrahedral site stabilizes the cubic spinel structure. Figure 2(b) - (g) illustrates the Rietveld refinement patterns of the RT - XRD patterns for all the samples. The refinement parameters (lattice parameters and atomic parameters) are provided in the supporting information file \cite{27}. All the samples, except NiCr$_2$O$_4$ crystallize in a cubic spinel structure (space group $Fd\overline{3}m$). The Chromium cations (Cr$^3$), located at the 16b (0.5, 0.5, 0.5) site do not show any deviation from the full occupancy. The refined lattice parameters, Cr - Cr and Cr - A bond distances (Figure 2 (h) and (i)) are found to increase with increasing the average radius of the A site. Figure 3 (a) - (e) displays the FE-SEM images along with the results of the elemental mapping for all the samples. The EDS mapping showcases the stoichiometric chemical composition, indicating the homogeneous distribution of all elements across the sample at a micrometer scale. The chemical compositions as obtained from the EDS analysis are highlighted in Table I. 

Furthermore, to investigate the detailed crystal structure, we have collected the neutron powder diffraction (NPD) patterns at room temperature (RT) for two high entropy spinel oxides [Ni$_{0.2}$Mn$_{0.2}$Co$_{0.2}$Cu$_{0.2}$Zn$_{0.2}$]Cr$_{2}$O$_{4}$, and [Ni$_{0.2}$Mg$_{0.2}$Co$_{0.2}$Cu$_{0.2}$Zn$_{0.2}$]Cr$_{2}$O$_{4}$. We will use the results of the NPD Rietveld refinements to describe the crystal structure of these two samples. Figure 4 shows the final plot of the RT-NPD Rietveld refinements for these two samples. The structural parameters obtained from the RT-NPD refinement for the [Ni$_{0.2}$Mn$_{0.2}$Co$_{0.2}$Cu$_{0.2}$Zn$_{0.2}$]Cr$_{2}$O$_{4}$, and [Ni$_{0.2}$Mg$_{0.2}$Co$_{0.2}$Cu$_{0.2}$Zn$_{0.2}$]Cr$_{2}$O$_{4}$ samples are summarized in Table II. 
Cr cations occupy the 16d (0.5, 0.5, 0.5) positions and form the octahedrally coordinated pyrochlore lattice of the spinel structure, while A site cations (Ni, Mn, Mg, Co, Cu, Zn) occupy the eightfold tetrahedral A sites (8a (0.125, 0.125, 0.125)). Oxygen atoms occupy the 32e (x, x, x) Wyckoff positions. Refinement of the oxygen and Cr occupancy indicates no deviation from full occupancy. In agreement with the EDS measurements, occupancy refinements of the A site elements for [Ni$_{0.2}$Mg$_{0.2}$Co$_{0.2}$Cu$_{0.2}$Zn$_{0.2}$]Cr$_{2}$O$_{4}$ shows the Mg deficiency in the A site. For [Ni$_{0.2}$Mn$_{0.2}$Co$_{0.2}$Cu$_{0.2}$Zn$_{0.2}$]Cr$_{2}$O$_{4}$ no significant deviations from the stoichiometric composition are observed in the refinement.  Fig. 5 presents the STEM-HAADF image of the [Ni$_{0.2}$Mg$_{0.2}$Co$_{0.2}$Cu$_{0.2}$Zn$_{0.2}$]Cr$_{2}$O$_{4}$ along the [101] direction. The observed STEM image corresponds well with the spinel material's anticipated crystal structure, highlighting the excellent structural quality at the nanoscale. The lattice parameter measured from the STEM images (8.4 \AA) is consistent with the values obtained from RT - NPD (Table II) and RT-XRD refinements.

\begin{figure}
\includegraphics[width=0.85\columnwidth]{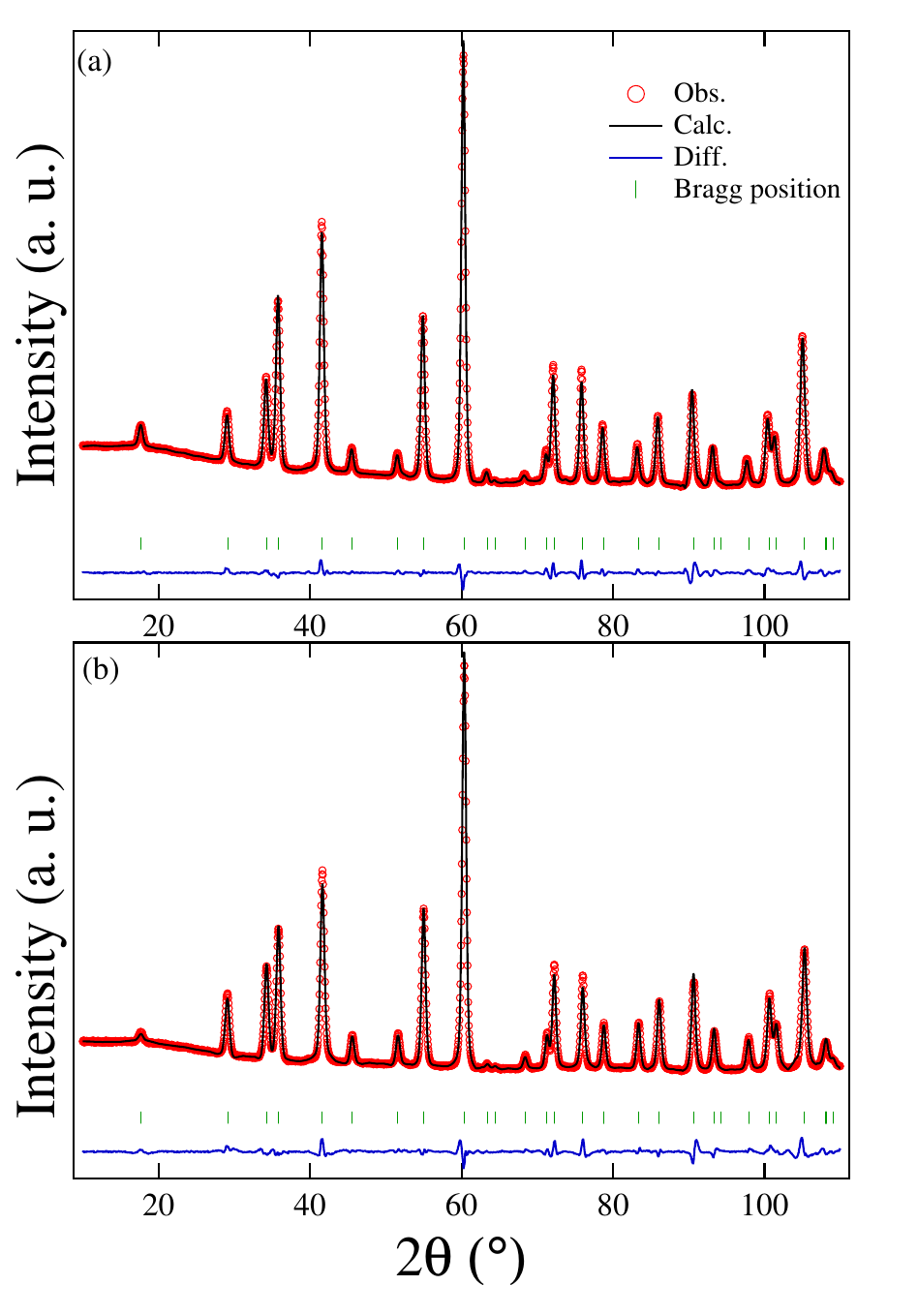}
\caption{\label{Figure 4:NPD} Rietveld refinement plots of the room temperature neutron powder diffraction (NPD) patterns for (a) [Ni$_{0.2}$Mn$_{0.2}$Co$_{0.2}$Cu$_{0.2}$Zn$_{0.2}$]Cr$_{2}$O$_{4}$ and (b) [Ni$_{0.2}$Mg$_{0.2}$Co$_{0.2}$Cu$_{0.2}$Zn$_{0.2}$]Cr$_{2}$O$_{4}$.}
\end{figure}

\begin{table*}
\small
  \caption{\ Atomic and structural parameters and selected bond lengths obtained from the RT - NPD refinements for [Ni$_{0.2}$Mn$_{0.2}$Co$_{0.2}$Cu$_{0.2}$Zn$_{0.2}$]Cr$_{2}$O$_{4}$ and [Ni$_{0.2}$Mg$_{0.2}$Co$_{0.2}$Cu$_{0.2}$Zn$_{0.2}$]Cr$_{2}$O$_{4}$.}
  \label{tbl 2:NPD}
  \begin{tabular*}{\textwidth}{@{\extracolsep{\fill}}lll}
    \hline
      & [Ni$_{0.2}$Mn$_{0.2}$Co$_{0.2}$Cu$_{0.2}$Zn$_{0.2}$]Cr$_{2}$O$_{4}$ & [Ni$_{0.2}$Mg$_{0.2}$Co$_{0.2}$Cu$_{0.2}$Zn$_{0.2}$]Cr$_{2}$O$_{4}$\\
    \hline
    Space Group & $Fd\overline{3}m$ & $Fd\overline{3}m$\\

    a = b = c (\AA) & 8.3377 (1) \AA  & 8.3235 (1) \AA\\
    Ni (0.125, 0.125, 0.125) &  & \\
    Occupancy & 0.2 & 0.21 (1)\\
    B$_{iso}$ & 0.95 (4) & 0.82 (3)\\
     Mn/Mg (0.125, 0.125, 0.125) &  & \\
    Occupancy & 0.2 & 0.15 (1)\\
    B$_{iso}$ & 0.95 (4) & 0.82 (3)\\
     Co (0.125, 0.125, 0.125) &  & \\
    Occupancy & 0.2 & 0.22 (1)\\
    B$_{iso}$ & 0.95 (4) & 0.82 (3)\\
     Cu (0.125, 0.125, 0.125) &  & \\
    Occupancy & 0.2 & 0.21 (1)\\
    B$_{iso}$ & 0.95 (4) & 0.82 (3)\\
     Zn (0.125, 0.125, 0.125) &  & \\
    Occupancy & 0.2 & 0.21 (1)\\
    B$_{iso}$ & 0.95 (4) & 0.82 (3)\\
     Cr (0.5, 0.5, 0.5) &  & \\
    Occupancy & 1.0 & 1.0\\
    B$_{iso}$ & 0.52 (3) & 0.21 (2)\\
     O (x,x,x) &  & \\
    x = 0.26224 (3) & 0.26205 (3)\\
    Occupancy & 1.0 & 1.0\\
    B$_{iso}$ & 0.36 (3) & 0.26 (1)\\
    Cr - Cr & 2.9478 (1) \AA & 2.9428 (1) \AA\\
    Cr - A & 3.4566 (1) \AA & 3.4508 (1) \AA\\
     $\chi^2$ & 8.04 & 7.46\\
      R$_{P}$ & 5.84 & 6.98 \\
        R$_{WP}$ & 5.37 & 6.5 \\
    \hline
  \end{tabular*}
\end{table*}

\begin{figure}
\includegraphics[width=1\columnwidth]{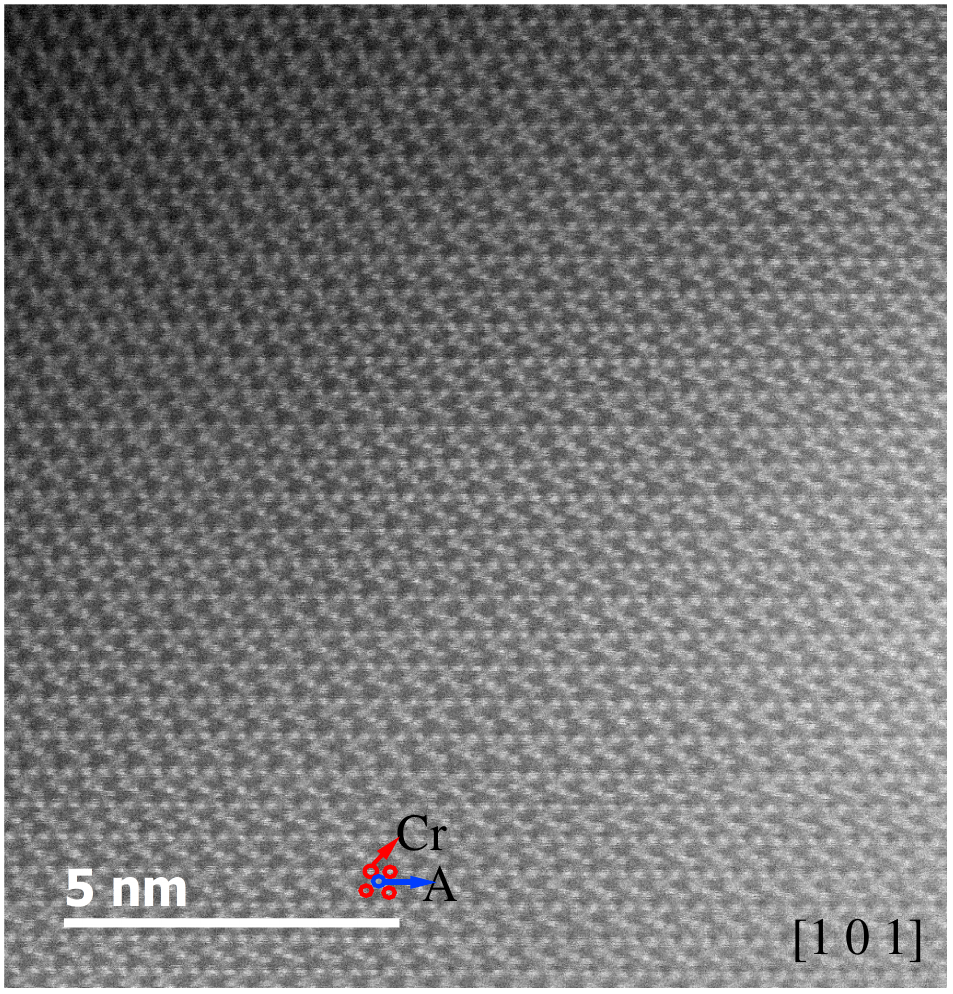}
\caption{\label{Figure 5:STEM} STEM - HAADF image along the [101] zone axis for [Ni$_{0.2}$Mg$_{0.2}$Co$_{0.2}$Cu$_{0.2}$Zn$_{0.2}$]Cr$_{2}$O$_{4}$, collected at room temperature. Cr and A site cations are shown in the figure.}
\end{figure}

\begin{figure}
\includegraphics[width=1.0\columnwidth]{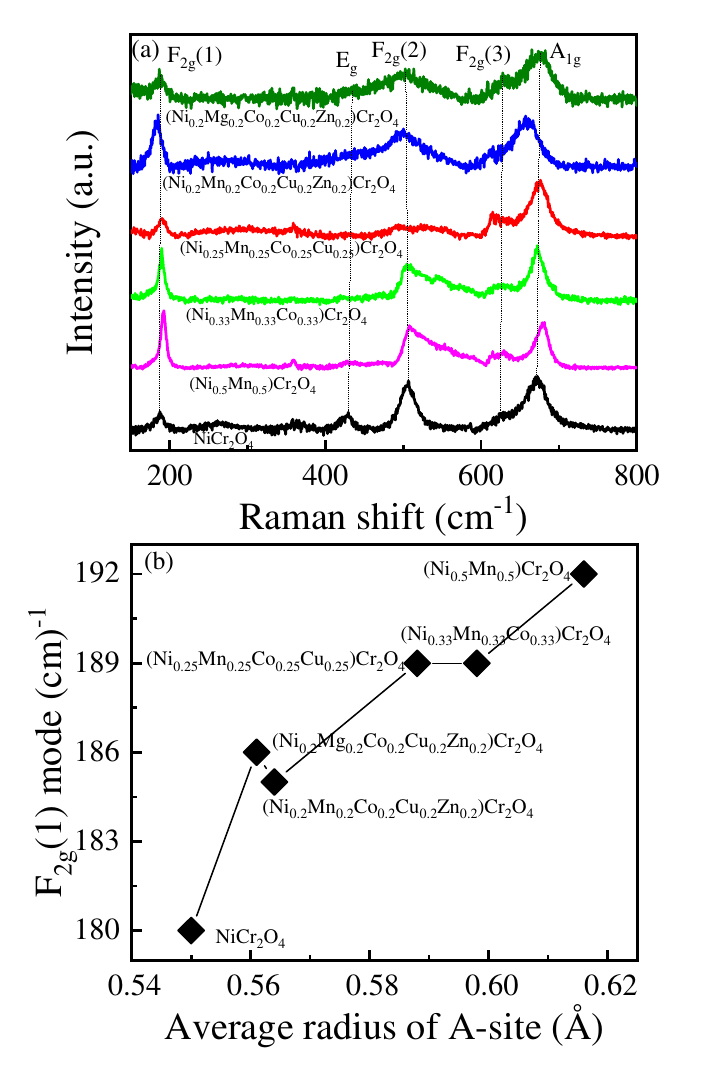}
\caption{\label{Figure 6:RAMAN} (a) Raman spectra for all the materials collected at room temperature. (b) Shift of the F$_{2g}$ (1) mode with the variation of the radius of the A site for all the samples.}
\end{figure}

\begin{figure}
\includegraphics[width=1.1\columnwidth]{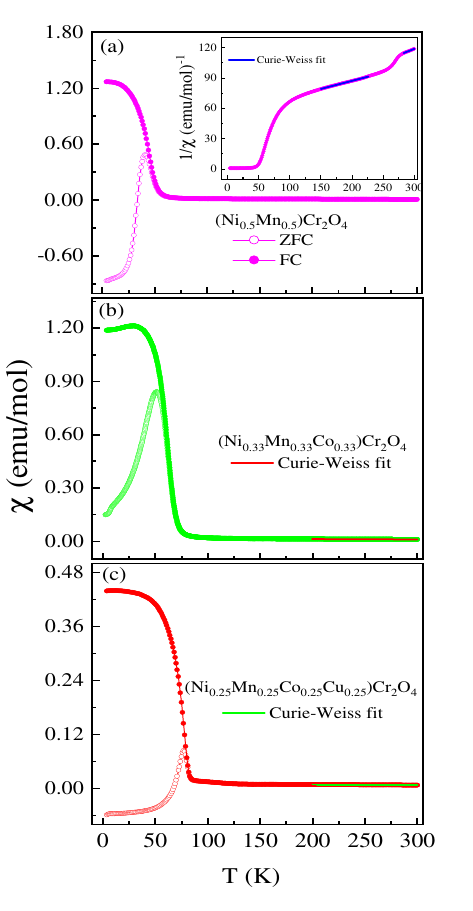}
\caption{\label{Figure 7:MTOe} Temperature dependent FC and ZFC magnetic susceptibilities for (a) [Ni$_{0.5}$Mn$_{0.5}$]Cr$_{2}$O$_{4}$, (b) [Ni$_{0.33}$Mn$_{0.33}$Co$_{0.33}$]Cr$_{2}$O$_{4}$, (c) [Ni$_{0.25}$Mn$_{0.25}$Co$_{0.25}$Cu$_{0.25}$]Cr$_{2}$O$_{4}$. The Curie-Weiss fitting for all the materials is highlighted in the graph. The inset in Figure (a) shows an anomaly due to the phase transition driven by Jahn-Teller distortions of the Ni$^2$ in the inverse susceptibility plot. The inclusion of Mn in the A-site of the structure dilutes the Jahn-Teller activity of the Ni cations. Therefore, the structural transition occurs at a lower temperature (transition temperature = 310 K for NiCr$_2$O$_4$).}
\end{figure}

\begin{figure}
\includegraphics[width=1.1\columnwidth]{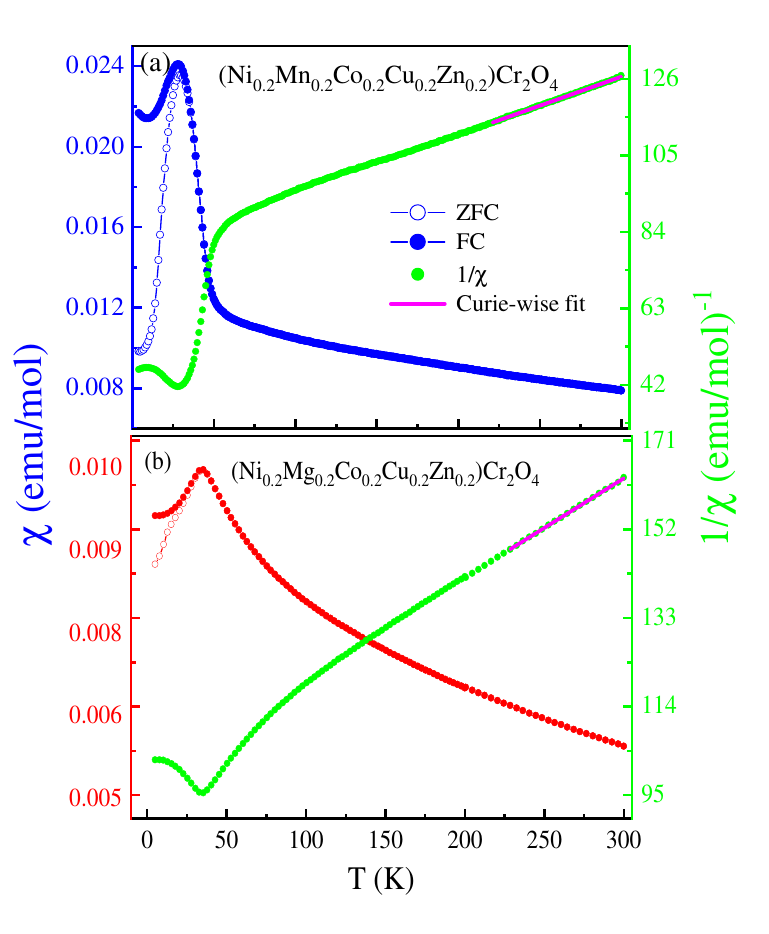}
\caption{\label{Figure 8:MT 1T} Temperature dependent (left axis) FC and ZFC magnetic susceptibilities and (right axis) inverse FC susceptibilities for (a) [Ni$_{0.2}$Mn$_{0.2}$Co$_{0.2}$Cu$_{0.2}$Zn$_{0.2}$]Cr$_{2}$O$_{4}$, and (b) [Ni$_{0.2}$Mg$_{0.2}$Co$_{0.2}$Cu$_{0.2}$Zn$_{0.2}$]Cr$_{2}$O$_{4}$.}
\end{figure}

\begin{figure}
\includegraphics[width=1.1\columnwidth]{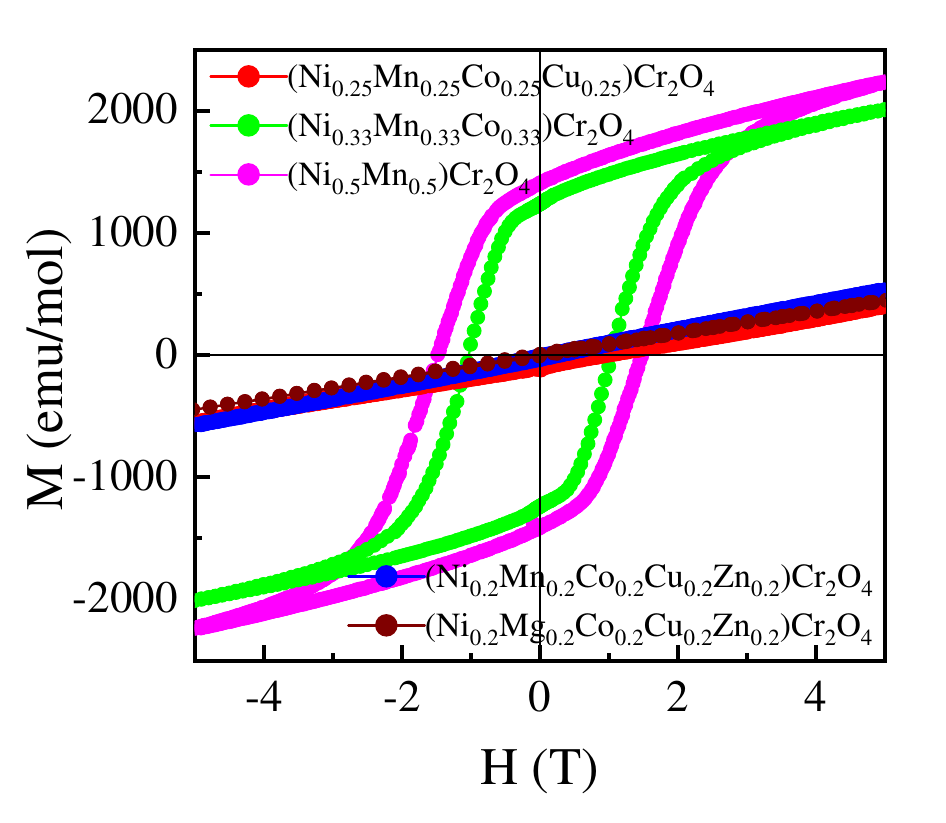}
\caption{\label{Figure 9:MH} Magnetic field variation of the magnetizations (M - H) recorded at 4 K for all the materials.}
\end{figure}

The Raman scattering studies were conducted at room temperature on all samples to understand their structure and the response of the increased disorder (and entropy) in the A site of these spinel materials. The room temperature Raman spectra for six different medium, and high entropy oxides (HEOs) with cubic structures are shown in Figure 6. The spinel unit cell, comprising 56 atoms, reduces to 14 atoms in the primitive cell. Based on group theory analysis, for Cubic spinels (space group Fd - 3m), five first-order Raman active phonons, represented as $\Gamma$$_{Raman}$ = A$_{1g}$ + E$_{g}$ + 3F$_{2g}$ are expected \cite{28, 29}. A$_{1g}$ mode (600-720 cm$^{-1}$) is due to vibrations (move away) along the direction connecting an oxygen atom to the tetrahedral A$^{2+}$ cations. The E$_{g}$ mode (250-360 cm$^{-1}$) corresponds to the symmetric bending of the oxygen anion relative to the octahedral B$^{3+}$ cations. The F$_{2g}$ (1) mode, appearing between 160-220 cm$^{-1}$, is related to a complete translation of the tetrahedral unit, while the higher frequency F$_{2g}$ (2) and F$_{2g}$ (3) modes (440-590 cm$^{-1}$) arise from asymmetric stretching and bending of oxygen, respectively. All five first-order Raman modes are apparent in all the samples. However, in comparison to NiCr$_{2}$O$_{4}$, we observe a broadening of the A$_{1g}$ peak in the medium and high entropy spinels. The electronic and high entropy stabilized disorder in the A site of the spinel structure could contribute to the broadening of the A$_{1g}$ peak. Figure 6 (b) highlights the evolution of the F$_{2g}$ (1) mode with the average radius of the A site (AB$_2$O$_4$ type) in the structure. It shows that the F$_{2g}$ (1) mode shifts toward the higher wave number with an increase in the average radius of the A site of the structure. Previously, it was suggested that F$_{2g}$ (1) mode corresponds to a translational motion of the entire A-site tetrahedral unit within the spinel structure \cite{28}. Therefore, our data support that this vibration primarily involves the tetrahedral cation, as this mode exhibits a significant and systematic change in the wavenumber among all Raman-active modes.

\begin{figure}
\includegraphics[width=1.0\columnwidth]{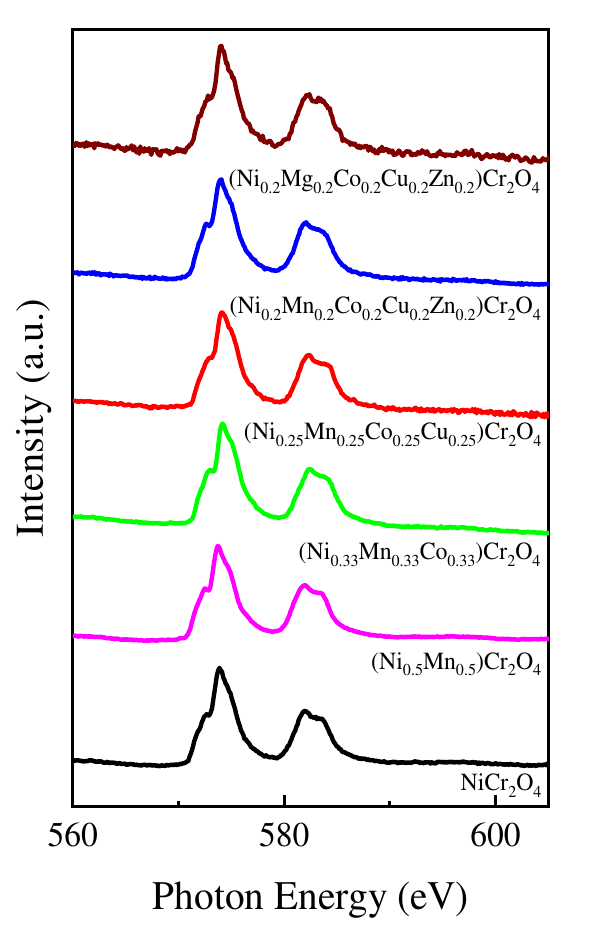}
\caption{\label{Figure 10:XAS} Room temperature XAS data of Cr L - edge for all the materials.}
\end{figure}

\begin{table*}
\small
  \caption{\ Sample compositions, magnetic transition temperatures (T$_C$ for ferrimagnetic (FIM) materials and T$_N$ for antiferromagnetic (AFM) materials), Coercivity (H$_C$), experimentally obtained and theoretically calculated effective magnetic moments ($\mu_{eff}$), Curie-Weiss temperature ($\theta_{CW}$) and frustration parameters (f = |$\frac{\theta_{CW}}{T_C}$|) for all the materials.}
  \label{tbl 3:Magnetic}
  \begin{tabular*}{\textwidth}{@{\extracolsep{\fill}}llllllll}
    \hline
    Sample Compositions & T$_{C}$ (K) & H$_{C}$ (4 K) & $\mu_{eff}$ ($\mu_{B}$) & $\mu_{eff}$ (theo.) & $\theta_{CW}$ & f & Ref.\\
    \hline
    NiCr$_{2}$O$_{4}$ & 74 (FIM) & 9kOe & 6.53 & 6.16$\mu_{B}$ & -487 & 6.6 & \cite{25}\\
    
   [Ni$_{0.5}$Mn$_{0.5}$]Cr$_{2}$O$_{4}$ & 45 (FIM) & 15kOe & 7.12 & 7.18$\mu_{B}$ & -351 & 7.8 & this work\\

   [Ni$_{0.33}$Mn$_{0.33}$Co$_{0.33}$]Cr$_{2}$O$_{4}$ & 63 (FIM) & 10.3kOe & 7.4 & 7.01$\mu_{B}$ & -375 & 5.9 & this work\\

   [Ni$_{0.25}$Mn$_{0.25}$Co$_{0.25}$Cu$_{0.25}$]Cr$_{2}$O$_{4}$ & 78 (FIM) & 0.7kOe & 6.86 & 6.72$\mu_{B}$ & -472 & 6 & this work\\

    [Ni$_{0.2}$Mn$_{0.2}$Co$_{0.2}$Cu$_{0.2}$Zn$_{0.2}$]Cr$_{2}$O$_{4}$ & 40 (FIM) & 0.17kOe & 6.74 & 6.49$\mu_{B}$ & -445 & 11.1 & this work\\

  [Ni$_{0.2}$Mg$_{0.2}$Co$_{0.2}$Cu$_{0.2}$Zn$_{0.2}$]Cr$_{2}$O$_{4}$ & 35 (AFM) & 0.08kOe & 6.13 & 5.93$\mu_{B}$ & -467 & 13.4 & this work, \cite{30}\\
    \hline
  \end{tabular*}
\end{table*}

Magnetic susceptibility as a function of temperature under field cooled (FC) and zero field cooled (ZFC) protocol are shown in Figure  \ref{Figure 7:MTOe} and \ref{Figure 8:MT 1T} for all the compositions. For medium entropy samples [Ni$_{0.5}$Mn$_{0.5}$]Cr$_{2}$O$_{4}$, [Ni$_{0.33}$Mn$_{0.33}$Co$_{0.33}$]Cr$_{2}$O$_{4}$, [Ni$_{0.25}$Mn$_{0.25}$Co$_{0.25}$Cu$_{0.25}$]Cr$_{2}$O$_{4}$, a sharp upturn is observed in the FC susceptibility data and is marked as the onset of the magnetic ordering. The corresponding transition temperatures for different samples are tabulated in Table 3.  Similar to the parent NiCr$_{2}$O$_{4}$ these three samples show ferrimagnetic (FIM) - type ordering. An anomaly in the susceptibility plot for [Ni$_{0.5}$Mn$_{0.5}$]Cr$_{2}$O$_{4}$ is observed at 280 K and could be an indication of tetragonal to cubic structural transition. In NiCr$_{2}$O$_{4}$ similar structural transition occurs at 310 K \cite{25}. At 310 K, a phase transition driven by cooperative Jahn-Teller distortions occurs and lifts the orbital degeneracy and reduces the structural symmetry from cubic (Fd-3m) to tetragonal (I4$_{1}$/amd). The inclusion of Mn in the A-site of the structure (in [Ni$_{0.5}$Mn$_{0.5}$]Cr$_{2}$O$_{4}$) dilutes the Jahn-Teller distortion of the Ni cations. Therefore, the structural transition is occurring at a lower temperature (at 280 K). Further dilution in the Jahn-Teller activity in other medium and high entropy samples removes the tetragonal distortion in the sample.  High irreversibility between the field-cooled and zero-field-cooled measurements is observed for these three samples and can be attributed to spin freezing. This spin freezing likely results from the high degree of structural disorder in the sample. The magnetic field-dependent magnetization measurements for all the samples recorded at 4 K are shown in Figure 9. The typical Brillouin shape of the curves for  (Ni$_{0.5}$Mn$_{0.5}$)Cr$_2$O$_4$, and (Ni$_{0.33}$Mn$_{0.33}$Co$_{0.33}$)Cr$_2$O$_4$ is a signature of the typical ferrimagnetic ordering of the samples. Also, an extremely high value of the coercivity is observed ($\approx$ 1.5 T, table 3) for these two samples. However, for [Ni$_{0.25}$Mn$_{0.25}$Co$_{0.25}$Cu$_{0.25}$]Cr$_{2}$O$_{4}$ a linear complex antiferromagnetic type M - H loop is observed. Spinel oxides based on Cr often exhibit intricate magnetic behaviors. For instance, CoCr$_2$O$_4$ transitions to a ferrimagnetic state at 95 K, followed by long-range non-collinear magnetic ordering below 31 K \cite{16, 31}. NiCr$_2$O$_4$ shows ferrimagnetic ordering at
74 K followed by a complex antiferromagnetic ordering
at 31 K \cite{25}. The linear M-H characteristic in [Ni$_{0.25}$Mn$_{0.25}$Co$_{0.25}$Cu$_{0.25}$]Cr$_{2}$O$_{4}$ could be due to predominant antiferromagnetic interactions overshadowing ferrimagnetic interactions at lower temperatures. The FC and ZFC magnetic susceptibilities as a function of temperature for two high entropy samples [Ni$_{0.2}$Mn$_{0.2}$Co$_{0.2}$Cu$_{0.2}$Zn$_{0.2}$]Cr$_{2}$O$_{4}$, and [Ni$_{0.2}$Mg$_{0.2}$Co$_{0.2}$Cu$_{0.2}$Zn$_{0.2}$]Cr$_{2}$O$_{4}$ are shown in Figure  \ref{Figure 8:MT 1T}. Similar to other samples both the FC and ZFC susceptibility for [Ni$_{0.2}$Mn$_{0.2}$Co$_{0.2}$Cu$_{0.2}$Zn$_{0.2}$]Cr$_{2}$O$_{4}$ show a ferrimagnetic type sharp upturn at 40 K, then on further cooling irreversibility between FC and ZFC susceptibility data is observed below 25 K. On the other hand [Ni$_{0.2}$Mg$_{0.2}$Co$_{0.2}$Cu$_{0.2}$Zn$_{0.2}$]Cr$_{2}$O$_{4}$ shows a long-range antiferromagnetic ordering at 35 K \cite{30}. To further investigate the magnetic properties of these samples, the temperature variation of the susceptibility plots for all the samples are fitted with the Curie-Weiss expression 
\begin{equation}
\chi (T) = \frac {C} {(T-\theta_{CW})}
\label{eqn1:SP}
\end{equation}
where $\theta_{CW}$ is the Curie-Weiss temperature, C is the Curie-Weiss constant, and the fitted parameters are listed in table III. The calculated effective magnetic moments (P$_{eff}$) using the fitted Curie-Weiss constant C show good agreement with the expected theoretical values (considering the spin-only contribution) for all the samples. However, all the materials show very high values of $\theta_{CW}$. All obtained Curie-Weiss temperatures ($\theta_{CW}$) are negative. This indicates that the overall exchange interactions among the magnetic ions are predominantly antiferromagnetic. In comparison with the low entropy and medium entropy samples two high entropy materials show a significant increase in the frustration parameter f = |$\frac{\theta_{CW}}{T_C}$|. The high atomic-level disorder in these two high entropy samples likely results in considerable magnetic frustration. The complex crystal structure, featuring diamond and pyrochlore lattices, along with the random distribution of magnetic and nonmagnetic cations within distinct sublattices, is anticipated to foster complex magnetic correlations and substantial frustration.

In general, the intricate magnetic properties observed in these samples reflect the presence of multiple competing magnetic interactions. This is expected in systems where multiple magnetic cations occupy different crystal field environments, resulting in varied magnetic moments that interact with diverse exchange interaction strengths and signs. The magnetic transition temperature is observed to vary with changes in the entropy of the structure, yet this variation does not follow a consistent increasing or decreasing trend. Since the transition temperature is dependent on the strength of exchange interactions, this suggests a complex interplay among the magnetic ions. This complexity precludes straightforward comparisons across systems with differing types and proportions of magnetic ions, preventing the establishment of a systematic trend.

To explore the electronic states and local geometry of the material, XAS experiments were performed on the high entropy samples. Figure 10 compares the XAS spectra of the Cr L$_{3,2}$-edges for all the compounds. In Cr-based spinel materials, Cr cations occupy the octahedral B site. Figure 10 clearly shows that all the Cr L$_{3,2}$ spectra are similar to those of the parent NiCr$_2$O$_4$, confirming the octahedral environment of the Cr$^{3+}$ cations in all the samples. Consequently, all the A-site cations in these samples are expected to have a charge state of 2+. 

\section{Conclusion}
In this study, we successfully synthesized and characterized a new series of Cr-based low, medium and high-entropy spinel oxides with composition NiCr$_{2}$O$_{4}$, [Ni$_{0.5}$Mn$_{0.5}$]Cr$_{2}$O$_{4}$, [Ni$_{0.33}$Mn$_{0.33}$Co$_{0.33}$]Cr$_{2}$O$_{4}$, [Ni$_{0.25}$Mn$_{0.25}$Co$_{0.25}$Cu$_{0.25}$]Cr$_{2}$O$_{4}$, [Ni$_{0.2}$Mn$_{0.2}$Co$_{0.2}$Cu$_{0.2}$Zn$_{0.2}$]Cr$_{2}$O$_{4}$, and [Ni$_{0.2}$Mg$_{0.2}$Co$_{0.2}$Cu$_{0.2}$Zn$_{0.2}$]Cr$_{2}$O$_{4}$. RT-XRD patterns and RT-EDS elemental mappings are used to determine the crystal structure and micro-scale homogeneity for all the samples. The STEM-HAADF image of the [Ni$_{0.2}$Mg$_{0.2}$Co$_{0.2}$Cu$_{0.2}$Zn$_{0.2}$]Cr$_{2}$O$_{4}$ highlight the excellent crystalline quality and homogeneity in nanoscale. The lattice parameters and Cr-Cr distances for all the samples are found to follow the average radius of the A site. Furthermore, we have used the ambient temperature neutron diffraction patterns to determine the detailed crystal structure for two high entropy spinel oxides [Ni$_{0.2}$Mn$_{0.2}$Co$_{0.2}$Cu$_{0.2}$Zn$_{0.2}$]Cr$_{2}$O$_{4}$, and [Ni$_{0.2}$Mg$_{0.2}$Co$_{0.2}$Cu$_{0.2}$Zn$_{0.2}$]Cr$_{2}$O$_{4}$. Magnetic measurements reveal that introducing multiple cations into the A-site of the spinel structure profoundly affects the magnetic properties, inducing a range of behaviors from ferrimagnetism to antiferromagnetism, and significantly enhancing coercivity. High entropy sample [Ni$_{0.2}$Mn$_{0.2}$Co$_{0.2}$Cu$_{0.2}$Zn$_{0.2}$]Cr$_{2}$O$_{4}$ shows a complex ferrimagnetic behaviour at 40 K. On the other hand, a small change in the structure induces antiferromagnetic behavior in [Ni$_{0.2}$Mg$_{0.2}$Co$_{0.2}$Cu$_{0.2}$Zn$_{0.2}$]Cr$_{2}$O$_{4}$. Extremely high coercivity is observed for [Ni$_{0.5}$Mn$_{0.5}$]Cr$_{2}$O$_{4}$, [Ni$_{0.33}$Mn$_{0.33}$Co$_{0.33}$]Cr$_{2}$O$_{4}$. Room temperature XAS confirms the octahedral environment of the Cr$^{3+}$ cations in all the samples.  The variation in magnetic transition temperatures, independent of a clear trend with entropy changes, indicates the sensitivity of magnetic properties to specific elemental compositions and their distributions within the lattice. These findings demonstrate the versatility and potential of high-entropy spinel oxides in developing materials with tailored magnetic properties for various technological applications. From a broader vantage point, our study highlights the potential to incorporate a spectrum of elemental components within a structurally complex oxide framework. This highlights excellent opportunities for tailoring and refining properties in strongly correlated materials.

\section{Acknowledgments}

S. M. acknowledges the SERB, Government of India, for the (SRG/2021/001993) Start-up Research Grant. We acknowledge the UGC-DAE Consortium for Scientific Research, Indore for the XAS and magnetization measurements. UGC-DAE Mumbai is also acknowledged for the Neutron diffraction experiments. We would like to thank Dr. Rajeev Rawat (magnetization), Dr. S. D. Kaushik (Neutron diffraction), Mr. Kranti Kumar Sharma (magnetization) and Rakesh Shah (SXAS) for their support during the measurements.


\begin{thebibliography}{References}

\bibitem{1} D. B. Miracle, O. N. Senkov, Acta Mater., 122, 448-511 (2017).
\bibitem{2} Y. F. Ye, Q. Wang, J. Lu, C. T. Liu, Y. Yang, Materials Today, 19, 349-362 (2016).
\bibitem{3} C. M. Rost, E. Sachet, T. Borman, A. Moballegh, E. C. Dickey, D Hou, J. L. Jones, S. Curtarolo, J.-P. Maria, Nat. Commun., 6, 8485 (2015).
\bibitem{4} A. R. Mazza, E. Skoropata, Y. Sharma, J. Lapano, T. W. Heitmann, B. L. Musico, V. Keppens, Z. Gai, J. W. Freeland, T. R. Charlton, M. Brahlek, A. Moreo, E. Dagotto and T. Z. Ward, Adv. Sci., 9, 2200391 (2022).
\bibitem{5} G. H. J. Johnstone, M. U. Gonzalez-Rivas, K. M. Taddei, R. G. A. Sawatzky, R. J. Green, M. Oudah, and A. M. Hallas, J. Am. Chem. Soc., 144, 20590-20600 (2022).
\bibitem{6} N. Sharma, S. Jangid, S. Choudhury, S. Kr Mahatha, R. P. Singh, S. Marik, Applied Physics Letters 123, 161901 (2023).
\bibitem{7} S Marik, D Singh, B Gonano, F Veillon, D Pelloquin, Y Breard, Scripta Mater., 186, 366-369 (2022).
\bibitem{8} M. Cocconcelli, D. Miertschin, B. Regmi, D. Crater, F. Stramaglia, L. Yao, R. Bertacco, C. Piamonteze, S. van Dijken, and A. Farhan, Phys. Rev. B, 109, 134422 (2024).
\bibitem{9} F. Jin, Y. Zhu, L. Li, Z. Pan, D. Pan, M. Gu, Q. Li, L. Chen, H. Wang, Advanced Functional Materials, 33, 2214273 (2023).
\bibitem{10} A. Sarkar, D. Wang, M. V. Kante, L. Eiselt, V. Trouillet, G. Iankevich, Z. Zhao, S. S. 
Bhattacharya, H. Hahn, R. Kruk, Advanced Materials, 35, 2207436 (2022).
\bibitem{11} R. R. Katzbaer, F. M. Vieira, I. Dabo, Z. Mao and R.E. Schaak, J. Am. 
Chem. Soc. 145, 6753-6761 (2023).
\bibitem{12} F. Cheng, Z. Meng, C. Cheng, J. Hou, Y. Liu, B. Ren, H. Hu, F. Gao, Y. Miao and X. Wang, Corrosion Science, 218, 111199 (2023).
\bibitem{13} S. S. Jana and T. Maiti, Mater. Horiz., 10, 1848-1855 (2023).
\bibitem{14} Na-Li Chen, Ge-Ting Sun, Cheng-Yu He, Bao-Hua Liu, Hui-Xia Feng, Gang Liu and Xiang-Hu Gao, Materials Today Physics, 42, 101363 (2024).
\bibitem{15} V. Fritsch, J. Hemberger, N. Buttgen, E.-W. Scheidt, H.-A. Krug von Nidda, A. Loidl, and V. Tsurkan, Phys. Rev. Lett., 92, 116401 (2004).
\bibitem{16} V. Kocsis, S. Bordács, D. Varjas, K. Penc, A. Abouelsayed, C. A. Kuntscher, K. Ohgushi, Y. Tokura, and I. Kezsmarki, Phys. Rev. B, 87, 064416 (2013).
\bibitem{17} R. Fichtl, V. Tsurkan, P. Lunkenheimer, J. Hemberger, V. Fritsch, H.-A. Krug von Nidda, E.-W. Scheidt, and A. Loidl, Phys. Rev. Lett. 94, 027601 (2005).
\bibitem{18} R. Nirmala, Kwang-Hyun Jang, H. Sim, H. Cho, J. Lee, Nam-Geun Yang, S. Lee, R. M. Ibberson, K. Kakurai and M. Matsuda, Journal of Physics: Condensed Matter, 29, 13LT01 (2017).
\bibitem{19}J. H. Chung, M. Matsuda, S.-H. Lee, K. Kakurai, H. Ueda, T. J. Sato, H. Takagi, K.-P. Hong, and S. Park, Phys. Rev. Lett. 95, 247204 (2005).
\bibitem{20} L. D. C. Jaubert, Y. Iqbal and Harald O. Jeschke, arXiv:2405.13908v1 (2024).
\bibitem{21} J. R. Chamorro, L. Ge, J. Flynn, M. A. Subramanian, M. Mourigal, and T. M. McQueen, Phys. Rev. Materials 2, 034404 (2018).
\bibitem{22} D. Bergman, J. Alicea, E. Gull, S. Trebst, and L. Balents, Nature Phys. 3, 487 (2007).
\bibitem{23} F. L. Buessen, M. Hering, J. Reuther, and S. Trebst, Phys. Rev. Lett. 120, 057201 (2018).
\bibitem{24} M. K. Wallace, Jun Li, P. G. Labarre, S. Svadlenak, D. Haskel, J. Kim, G. E. Sterbinsky, F. Rodolakis, H. Park, A. P. Ramirez, and M. A. Subramanian, Phys. Rev. Materials 5, 094410 (2021).
\bibitem{25} M. R. Suchomel, D. P. Shoemaker, L. Ribaud, M. C. Kemei and R. Seshadri, Phys. Rev. B, 86, 054406 (2012).
\bibitem{26} O. Crottaz, F. Kubel and H. Schmid, J. Mater. Chem., 7, 143-146 (1997).
\bibitem{27} Supporting
\bibitem{28} V. D'Ippolito, G. B. Andreozzi, D. Bersani, P. P. Lottici, Journal of Raman Spectroscopy, 46, 1255--1264 (2015).
\bibitem{29} Z. Wang, S.K Saxena, P. Lazor and H.S.C O'Neill, Journal of Physics and Chemistry of Solids, 64, 425-431 (2003).
\bibitem{30} S Marik, D Singh, B Gonano, F Veillon, D Pelloquin and Y Breard, Scripta Materialia, 183, 107-110 (2020).
\bibitem{31} D. Zakutna, A. Alemayehu, J. Vlcek, K. Nemkovski, C. P. Grams, D. Niznansky, D. Honecker and S. Disch, Phys. Rev. B 100, 184427 (2019).



\end{thebibliography}
\end{document}